\documentclass[aps,superscriptaddress,onecolumn,amssymb,eqsecnum,showpacs,nofootinbib,showkeys]{revtex4}
\usepackage{graphicx}
\usepackage{bm}
\usepackage{amsmath}
\usepackage[latin1]{inputenc}
\usepackage[english]{babel}
\usepackage[T1]{fontenc}
\usepackage{amssymb}
\usepackage{amsfonts}
\usepackage{epsfig}
\usepackage{colordvi}
\usepackage{psfrag}
\usepackage{color}
\newcommand{\beq}{\begin{equation}}
\newcommand{\eneq}{\end{equation}}

\begin{document}

\title{ASPECTS OF DUALITY IN COSMOLOGY}

\author{Gabriele Gionti S. J.}
\email{ggionti@as.arizona.edu}
\address{Specola Vaticana, Vatican City, V-00120, Vatican City State and Vatican Observatory Research Group, Steward Observatory, The University Of Arizona, 933 North Cherry Avenue Tucson, Arizona 85721, USA}
\address{Istituto Nazionale di Fisica Nucleare (INFN), Laboratori Nazionali di Frascati, Via E. Fermi 40, 00044 Frascati, Italy}

\begin{abstract}
In the first part of this article, given the intent to stay at a popular level, it has been
introduced and explained briefly basic concepts of Einstein's General Relativity, 
Dark Matter, Dark Energy, String Theory, Quantum Gravity and Extended Theories of Gravity. 
The core of this research is based on selecting a class of $f(R)$ theories of gravity, which exhibits scale factor duality transformations. The starting point of this theory is the effective theory of gravity derived from Bosonic String Theory, which is called tree level effective theory of gravity. It is shown that this theory can be cast in a class of $f(R)$ theories of gravity (modified theories of Einstein's General Relativity). It is imposed that FLRW metric be solution of this class of $f(R)$ theories, and, using the Noether symmetry approach, it is found that the cosmological model has scale factor duality like the Pre-Big Bang cosmology of Gasperini and Veneziano.
\end{abstract}

\keywords{ Modified theories of gravity; string theory;  scale factor duality; conservation laws.}

\maketitle

\section{Introduction}

This essay starts with a brief description of what is General Relativity. The basic concepts of General Relativity, the concept of Space-Time, the principles of equivalence and general covariance are introduced and discussed. It is also emphasized that Einstein's General Relativity was born by a fundamental question of how one can formulate a theory of gravitation, which is not "an action at distance" but a field theory on the example of electromagnetism. Then, in Einstein's General Relativity,  Space-Time becomes a physical entity, in the sense that it is deformed by the mass-energy of bodies. 

The theory of Einstein's General Relativity has equations of evolution in which the quantity to determinate is the topology and the metric structure of Space-Time, which, in simple words, is the shape of the space and time in which one lives (for example the spatial shape of our Universe is the surface of a three dimensional sphere and the time is a semi-line, a line with an origin). Said in other words, solutions of Einstein's Equations are cosmological models of the Universe like the Friedmann , Lemaitre, Robertson, Walker (FLRW) cosmological model (which is isotropic and homogeneous in the spatial dimensions). FLRW cosmology is called the standard model of cosmology but, in order to fit with experimental data, needs some extra ingredients which are Dark Matter and Dark Energy that together FLRW cosmology is know as $\Lambda$CDM model. Here $\Lambda$  is the cosmological constant, which is introduced in the Einstein's Equations in order to include dark energy, and CDM stands for Cold Dark Matter. In fact, after many years of investigation it has been found that the dark matter known in the universe is at non relativistic energy (therefore it is "cold"). 

The main reason for introducing Dark Matter is due to the fact that, as it will be explained in the following, it is impossible to explain the distribution of the radial velocities of spiral galaxies without introducing the hypothesis of an existence of some kind of not visible matter. On the other side, the existence of Dark Energy is needed to explain the recent observed acceleration of the universe. 

In this context, it is introduced the theory of extended theories of gravity (or alternative theories of gravity or f(R) theories), which are a generalization of Einstein's General Relativity and one of their aims is also to fit the recent cosmological data. In this respect, it is also mentioned the tests that the Planck collaboration has done to verify the validity of alterative theories of gravity. 

The core of this work remains the effort of showing that the tree level effective theory of gravity derived from bosonic string theory can be written as a class of $f(R)$ theories of gravity. For this purpose, it has been explained what is String Theory and, in general, what is a Quantum Theory of Gravity. Then, it is highlighted that Bosonic String Theory, at low energy, when the gravitational interaction dominates, has a behavior like a Brans-Dicke gravity with a scalar field that is called dilaton. It is shown that this type of theory can be cast, using Weyl's transformations and a hypothesis on the relation between the Weyl's mode and  dilaton field, in a class of $f(R)$ theories. 

In this context, it has been stressed the issue of Duality in String Theory. In general, duality is, in string theory, a symmetry that the solutions of the equations of motions have for which two different theories (two theories with different Lagrangian functions) have the same solutions. In particular, duality symmetry on the tree level effective theory of gravity, which derives from string theory, is a symmetry both of the theory (the Lagrangian function) and of the equations of motions and this generates cosmological models, which are called pre-Big Bang models or Gasperini-Veneziano models.

In casting a tree level effective theory of gravity derived from bosonic string theory into a $f(R)$ theory, it has been used the Noether symmetry approach. This is a method, which fixes appropriate mathematical relations under which a $f(R)$ theory of gravity has a Noether symmetry, which simplifies the problem. Substituting this conditions into the class of $f(R)$ theories of gravity, it has been shown that it is possible to have that the class of $f(R)$ theories gets a duality symmetry, in the sense that there exists a duality symmetry of the type a(t) $\mapsto$ 1/a(t). It is straightforward to recognize that this symmetry is a scale factor a(t) inversion as in the Gasperini-Veneziano model.

\section{General Relativity}

It is very nice coincidence that the 80th Vatican Observatory Anniversary symposium happens in the same year of the golden jubilee of Einstein's General Relativity. Before Einstein's General Relativity, the gravitational force was described by Newton's Universal Gravitational Law. The universal gravitational law was "an action at distance", in the sense that if there were two massive bodies, the perturbation of the position of one of the two bodies will affect the second body "immediately", independently by their relative distance. This implies that the signal of this perturbation will propagate with an infinite velocity. The discovery of the electromagnetic field and the fact that its perturbations propagate with a finite velocity, that's the light velocity, raised two questions: could the gravitational force be described by a field with a finite velocity of propagation?

The answer to this question was found by Einstein's General Theory of Relativity. He found a theory of the gravitational field that is a field theory. In Einstein's General Relativity (from now on G.R.) the perturbations of the Gravitational Field propagates with the velocity of light. G.R. is an extension of Special Relativity. Special Relativity is a theory in which all the physical phenomena are described by physical laws, which are invariant in form (the same mathematical expressions) if one passes from one reference frame to another reference frame that moves with a constant velocity respect to the first one (this class of reference frames are called "inertial reference frames"). In this class of reference frames, the velocity of light is, by axiom, the same in every reference frame. In General Relativity the physical phenomena are described by physical laws, which are invariant ("covariant" as said in the jargon of G.R., which means that their mathematical expressions are the same) for transformations from one reference frame to another reference frame, which has an arbitrary velocity respect to the first one (non inertial reference frames). The velocity of light is always the same also in non-inertial reference frames.

Summarizing the previous consideration, one can finally state that Einstein's General Relativity is a field theory of gravity, which is based on two principles: the principle of mass equivalence and the principle of covariance. The principle of mass equivalence says that gravitational mass, the property for which two bodies attract each other by the gravitational force, is the same quantity for which a body oppose resistance to the motion, property that is in called "inertia". That is gravitational mass and inertial mass are the same thing and have the same numerical value.

The second principle of G.R. is the covariance principle. This principle says that the laws of physics are the same (this property is called covariance and means that the laws of physics have the same mathematical expression) in every arbitrary reference frame. 

Space-Time becomes a physical entity, that is a dynamical quantity which is modified by the presence of massive bodies as well as (since the equivalence between mass and energy) the presence of energy. This is one of the greatest difference between General Relativity and any other field theory in physics. In fact, in any other theory field theory, Space-Time is like a box in which all the physical phenomena happens without any influence on it. On the contrary massive bodies and energy, in General Relativity, modifies Space-Time by generating a curvature on it, located around them. Free falling bodies move along the analogous of straight lines on the Space-Time surface, which are called geodedic lines. A geodedic line, by definition, is the line which makes stationary the functional of the distance between two points on the Space-Time surface. 

In this way, any perturbation of mass-energy propagates through Space-Time, which acts like a medium. This propagation happens always, by construction, at the velocity of light. Therefore General Relativity is a field theory of the Gravitational Field.  

\subsection{Einstein's Equations}

The Gravitational Field, in analogy with the Electromagnetic Field, is described by Field Equations. As it is usually done in Field Theory, Field Equations are derived by a variational principle, which, for Einstein's theory of Gravitation, consists in finding the stationary points of the action functional $S[M,g]$. Here $S[M,g]$ is a functional on a Lorentian Manifold $(M,g)$, $M$ being a Manifold and $g$ a Lorentian metric on it,  

\begin{equation}
S[M,g]=\int_{M}R\sqrt{-g}d^{4}x
\label{1}
\end {equation}

$R$ being the trace of the Ricci's tensor, which is a function of g.
The variation of $S$ gets the Einstein's Equation in the vacuum (without matter) 

\begin{equation}
R_{\mu\nu}-{1\over 2}g_{\mu\nu}R=0
\label{2}
\end {equation}

These field equations are six independent equations and the metric tensor $g$ is the quantity to determinate. The metric tensor is a rank two (roughly speaking a matrix) covariant tensor, which is a function of space-time. Once it is determined the metric tensor, the topology of Space-Time can be determined via the definition of open and closed sets through the metric structure.

Einstein's equations are six independent equations, four of which are first order equations in the time derivatives. Therefore, they are not equations which give a dynamical evolution of the system (Dynamical evolution is given by equations that have the second time derivative of the quantity to determinate). They are constraint equations, that is limitations on the possible  on the Cauchy's data (initial data) of the system. 

The number of equations which generate the evolution of the dynamical system is, finally, two and it is the reason for which the graviton particle, that is also the field associated at fundamental level to the gravitational interaction, has two degrees of freedom. 
 
 \section{Dark Matter and Dark Energy}
In recent years, astronomers have noticed that, studying the distribution of the radial velocity of matter as function of the distance from the center of galaxies, the mass observed does not explain the velocity distribution. The measurements of the radial velocity distribution can be explained well if one makes the hypothesis that there exists more matter than the observed one. This extra matter has been called Dark Matter, because it is not visible in the optical range \cite{Peebles1993}. This matter does not interact with radiation and it does not loose sufficiently kinetic energy to relax into the disk of galaxies, as it does baryonic matter (matter which contains nuclear particles, leptons and neutrons). This implies that this sort of matter is electrically neutral and it has been found that its velocity is far from being relativistic, therefore it is called {\it Cold}-Dark Matter.   

The study of a double galaxy cluster 1E0657-558 (the "bullet cluster") has confirmed the existence of cold Dark Matter, which has only gravitational interaction with itself and with baryonic matter \cite{Weinberg2008}. The amount of Dark Matter in the universe amounts at 26,8 of the total matter according to the recent measurements of the Planck satellite \cite{Planck2015resultsXIII} \cite{Planck2015resultsXIV}. Dark Matter behaves, gravitationally, as completely ordinary matter; for example, it causes gravitational lensing \cite{Weinberg2008}. 

Soon after the discovery of General Relativity, Friedmann-Lemaitre-Robertson-and Walker (FLRW) found a homogeneous and isotropic solution of Einstein's General Relativity which implied that the Universe was expanding. Einstein did not believe that this solution has a physical meaning, therefore introduced, in his equations, a so called "Cosmological constant", which kept the universe stationary (not expanding). Later, in 1929, it was discovered that all spectra of the stars exhibit a "Red Shift" of their absorption lines, which was a proof that the Universe was expanding. Then Einstein said that he did the worst mistake of his life \cite{Wald1984}.

In the early eighties of the last century, this mechanism of the cosmological constant was reconsidered for explaining an early period of the Universe in which it was thought to have expanded very fast, with an exponential rate, such that it could make sense to consider that all the different parts of the Cosmic Microwave Background Radiation were originally at thermal equilibrium. Cosmic Microwave Background Radiation is the "first light" from the Universe that was released 380 million of years after the Big Bang \cite{Peebles1993}. Before, the Universe was too small and every photon emitted was reabsorbed. Without invoking the mechanism of "Cosmological Inflation", FLRW metric solution of Einstein's Equations has a "particle Horizon", which means that not all the different regions of the Universe could, originally, be causally connected and then at thermal equilibrium.   
Cosmological Inflation, with an exponential expansion, implies an acceleration opposite to the gravitational attraction. This phenomenon is believed to be caused by the quantum vacuum of a field called the "Inflation". This vacuum has generated a huge contribution in the Einstein's General Relativity Equations driven by the cosmological constant, the same that before Einstein introduced and after withdraw when the galaxies' redshift and then the expansion of the universe was discovered \cite{Wald1984}. This produces in the equations of motions, in which one imposes the FLRW metric solution with matter that behaves as a fluid, a negative pressure, which gives reason to the emergence of a sort of anti-gravity force\cite{Weinberg2008}. This anti-gravity force explains a primordial huge expansion.

In 1998 observations of Type Ia supernovae showed that the Universe is expanding accelerating \cite{Weinberg2008}. This result implies, as in the case of Cosmic Inflation, that there exists a kind of force, which is opposite to Gravity. The mechanism known, up to now, to explain this antigravity force is "Dark Energy". Dark Energy is believed to be the vacuum point (zero point) energy of fundamental quantum fields. In fact, the zero-point energy of the harmonic oscillator, which is, basically, the system on which any quantum filed theory is based, is not zero but has a finite value. The sum of all these vacuum modes, for each fundamental interaction, generates an energy ("Dark Energy" because one does not see it), which is present in the Einstein's Equations via the cosmological constant  term. The last cosmological measures, obtained through the Planck satellite, seem to point out that the best theory capable to explain the cosmological measures is the $\Lambda$CDM model, 
that is the Einstein's theory of General Relativity with Cold Dark Matter and Dark Energy. In the $\Lambda$CDM model, the amount of Dark Energy is 68,3  \cite{Planck2015resultsXIV} and Einstein's General Relativity still remains the theory that explains better the cosmological data than other theories of gravity.

\section{Extended Theories of Gravity}

Extended theories of Gravity are "generalization" of Einstein's General Relativity. The principles, on which Extended theories of Gravity are based on, are the same of Einstein's General Relativity. The only, main, difference concerns the dynamics in the sense that their "action" and the relative equations of motions are, in general, respectively, generic functions of the trace of the Ricci tensor and equations of motion with higher order, more than two, derivatives of the metric tensor.   

Modifications to Einstein's General Relativity has been proposed since 1970 (see \cite{Sotiriou&Faraoni2010}) but mainly with the beginnings of the theory of cosmological inflation \cite{Starobinsky1980} it was clear that modifications of Einstein's General Relativity were important to understand the early period of our Universe. One of the "dreams" of extended theories of gravity has been to fit the experimental data, as showed in \cite{Capozziello&DeLaurentis2012}, with extended theories of gravity without introducing dark matter. But these models have too many parameters and appear quite phenomenological (see \cite{Capozziello&DeLaurentis2012}). Recent work of the Planck Collaboration \cite{Planck2015resultsXIV} have tried to study, on the base of recent data collected by the Planck satellite, how well the experimental data fit with modified theories of gravity. 

Extended theories of gravity are important also since, in many cases, they are the low energy limit of fundamental theories of Quantum Gravity. This, in simpler words, means that Extended theories of gravity can be considered, when the energies are not so high as in fundamental theories, a good approximation of the physical laws that regulates the quantum world in the very early times of our Universe. For example, String Theory, which is an attempt to find the quantum mechanical behavior of the gravitational field, in the limit of lower energies, when the gravitational interaction is supposed dominating over the other interactions, behaves as an Extended theory of gravity \cite{Tseytlin1991}.

One serious problem of extended theories of gravity is their quantization. In fact, once quantized, extended theories of gravity suffer lack of unitarity \cite{Capozziello&DeLaurentis2011}, which means, at quantum mechanical level, in analogy with classical mechanics, that if one changes the reference frame, Physics is affected by this change in the sense that the physical laws in the new reference frame appear different. This is one of the reasons for which people say that they are phenomenological theories and not fundamental, that is to say that they are "ad hoc" theories that works at same scales of energies but not at higher energies.   

An important aspect of alternative theories of gravity is the result (see \cite{DeFelice&Tsujikawa2010}) that $f(R)$ theories are equivalent to Brans-Dicke theory. This is a theory which, as concern its Lagrangian function, has the following characteristic: the trace of the Ricci tensor is multiplied with a scalar field; this scalar field has a potential and a kinetic term. It can be also shown that if one implements Weyl transformations on the metric tensor and choose a particular Weyl transformation, which is field dependent, \cite{DeFelice&Tsujikawa2010} a $f(R)$ theory can be shown to be equivalent to Einstein's Gravity with a potential and a kinetic term related to the scalar field.     

\section{String Theory}

String Theory is one of the proposal for a Quantum Theory of Gravity (Quantum Gravity). As it is quite well known, Einstein's equations have a singularity around the point t=0. This means that the physics, one knows, does not work around this initial point. The Physics that should describe our Universe around the point t=0 is called Quantum Gravity. A theory of Quantum Gravity is not known yet, there are many attempts to define a theory of Quantum Gravity, but no one can say that there exists a definitive formulation of Quantum Gravity (see \cite{Rovelli2004}). There exist many proposals for a Quantum Theory of Gravity. They can be divided into two main principal lines. A line that makes the assumption that the quantization of the gravitational field has to be performed through the quantization of Einstein's General Relativity, therefore without including all other fundamental interactions, and a line that promotes that the quantizzation of the gravitational field through the unification of all fundamental forces  at the energy for which (it is called Planck Energy) the laws of the gravitational field should behave according to Quantum Mechanics. As it is well known, Quantum Mechanics describes the laws of physics at microscopic level, that is at atomic and subatomic levels. Apart from gravity, the others fundamental interactions are: the Electromagnetic, the Weak Nuclear and the Strong Nuclear interaction. The Electromagnetic force regards the interaction among particles that have an electric charge. The Weak Nuclear force regards subatomic particles, which interact "weakly" (respect to the strong interaction), whose interactions are described by particle mediators which are massive. The Strong Nuclear force deals with  particles that interact "strongly" and are called Adrons. 

String Theory is a theory born at the end of the 60ties of the past century and it is considered a theory of Quantum Gravity. It is based on the "principle" that all fundamental forces in nature, the Electromagnetic Force, the Strong Nuclear Force, the Weak Nuclear Force and the Gravitational Force can be unified through a unique object that is called String. Particles are not, in this approach, the last building blocks of matter, but they are considered composite objects made out of Strings. The quantum oscillations of Strings will generate particles in the way one knows. Strings are one dimensional objects that could be open (Open Strings) and closed (Closed Strings). First excited level of Strings shows that their spectra contain particles of all the fundamental interactions. In this way the String, having in its first excited states all different particles of different interactions, realizes the unification of all the fundamental forces \cite{Greenetal1986} one knows in nature. 

Strings , in order to make sense mathematically, have to be embedded into a background space, which, for bosonic String Theory, is 26-dimensional. Bosonic String Theory is a theory that deals only with particles with integer spin (bosons). In fact, it happens that bosonic string theory has an "anomaly", which is called conformal anomaly. Anomaly means that a classical symmetry is not conserved at quantum level. Conformal Anomaly is a behaviour of the String Theory for which the classical theory is invariant under some specific transformations of the metric tensor that are called "conformal transformation", while the correspondent quantum theory results not invariant under conformal transformations. Therefore, in order to make the theory conformal invariant also at quantum level, one finds that has to embed  the string into a background space of 26 dimensions. People have considered also Supersymmetric String Theory, which is String Theory in which there is a further symmetry, which is called supersymmetry. This symmetry implies that it is possible to exchange particles with integer spin (Boson) with particle with semi-integer spin (Fermion) or, said in a different way, for ever integer spin particle there is a semi-integer spin partner particle. Supersymmetric Strings need to be embedded in a 10-dimensional space in order to avoid the conformal anomaly.     

At one loop, that means at the first order of approximation, at quantum level, in the case of Bosonic String Theory, the fact that the conformal symmetry has to be preserved implies that the trace of the Ricci's tensor plus the square of the first (covariant) derivatives of the scalar field, plus the D'Alambertian of the scalar field, plus the square of the derivatives of a field which is called B-field, plus a constant, in bosonic string theory in non critical dimensions (=different from 26 dimensions), have to be zero. This is considered an argument in favor of the fact that the Gravitational interaction is contained in Bosonic String Theory \cite{Greenetal1986}. 

One important feature of Bosonic String Theory is that it exibits a symmetry which is called "Duality". If the Lagrangian function of Bosonic String Theory has a particular symmetry, which means that there exists a vector field, an infinitesimal transformation, for which the Lagrangian function results to be invariant, then it is possible to define "dual picture", dual fields that determine a new Lagrangian function, whose equations of motion have the same solutions of the starting Lagrangian. This is an example of Duality and in particular is called Busher's Duality. In general, a scalar field called the "dilaton" is included in the mathematical formulation of Bosonic String Theory. The mathematical consequence of the preservation of the conformal invariance, as already said, implies that Einstein's equations with cosmological constant (in the case of non-critical string, that is the background Space-Time, in which the String is embedded, is not 26 dimensional) and a scalar field is satisfied. It can be eaisly shown that these equations derive from an action, that is the Einstein-Hilbert 's action plus a cosmological constant, plus a kinetic part of the scalar field (the dilaton) everything multiplied by the exponential of the dilaton field, which is a Brans-Dicke theory like (see for all details \cite{Capozzielloetal2016}\cite{me}). Generally speaking, the Lagrangian function could have more than one symmetry. If one has a theory which has several symmetries, in particular n independent symmetries which, in mathematical term, is said n independent and commuting isometries, then it is possible (see \cite{delaOssa&QuevedoF1993}) to define a dual theory such that the original theory and its dual, put together, are invariant under particular sets of transformations, which are called the group O(n,n). 

Gasperini and Veneziano \cite{Gasperini&Veneziano2003}  have studied into great details the cosmological models of the gravitational field which derives from bosonic string theory. As it has been extensively said, this is a Brans-Dicke like gravitational field with a kinetic term in the scalar field (dilaton), with the presence of the derivatives of a "B-field" (which is usually put to zero) and a cosmological constant in the case of non critical dimensions. This theory is manifestly O(n,n) invariant \cite{Meissner&Veneziano1991}, therefore one finds that, imposing that the theory has flat FLRW metric as solution of the equations of motions, there exists Duality symmetry transformations of the action and of the equations of motions \cite{Tseytlin&Vafa1992}. This theory is at the base of the Pre-Big Bang cosmology scenario in which there exists a phase of our universe before the time t=0. 

\section{From String Theory to Modified Gravity}

Suppose that in Bosonic String theory the gravitational interaction is dominating over the other interactions. In this case one knows, of course, that the graviton mode represents the main interaction. In this hypothesis, in the low energy limit at tree level Bosonic String theory becomes a Brans-Dicke-like gravity theory coupled to the dilaton field and with a B-field (a field which emerges in the low energy limit of Bosonic String Theory) and a cosmological constant (in the case in which one works in non critical dimensions, n). One question is if this theory can be cast in a $f(R)$ theory of gravity. One of the tracks followed has been to find a Weyl transformation on the metric tensor in such a way that the action of gravity, which derives from string theory, can be re-written as an alternative theory of gravity, a $f(R)$ theory (see \cite{Capozzielloetal2016}\cite{me}). A Weyl transformation is a transformation of the metric tensor, in which the metric tensor is equivalent to another metric tensor, in a different reference frame, multiplied by a generic scalar function (the Weyl's factor). 

The equivalence, in practice, is get by equating the integrands of the actions of the string based effective theory of gravity (with $B=0$) and the $f(R)$ theory. Using some results from the equations of motion and imposing that the Weyl's factor is linked to the dilaton field, it is possible to pin down a functional form for a class of $f(R)$ functions, which derive from bosonic string theory.  

Next step of this research is to study the cosmological consequences of this particular classes of $f(R)$ functions. For this reason, one consider a FLRW metric in the flat case. FLRW universe is like, in two dimensions, the surface of a sphere which is expanding and its radius measures the time, since the Big Bang, of the universe. In particular, one, for simplifying the problem, restricts to the flat case in which the space is flat and the distance between points in the space increases with time. 

Imposing that this metric be solution of these $f(R)$ theories, one obtains a Lagrangian  function, which depends also from the scale factor of the Universe.   

Given this particular classes of $f(R)$ functions, following a general method, one implements the FLRW metric solution introducing a new action functional in which there is the $f(R)$ function plus a Lagrange multiplier which constrains the trace of the Ricci curvature to be equal to the Ricci curvature evaluated   on FLRW metric.  

At this point, in order to make the theory mathematically  simpler, one uses a technique quite useful in many cases of study for $f(R)$ theories of gravity, which is called Noether symmetry approach. Noether symmetry approach is a procedure, which studies the general properties that a $f(R)$ theory of gravity have in order that its Lagrangian function has a symmetry. In general, a physical system may have some type of symmetries, like the translational symmetry, the rotational symmetry etc. These kinds of symmetries become groups of transformations. They are symmetries of the original Lagrangian function and of the solutions of its related equations of motion. In fact, these transformations map solutions of these equations of motion into new solutions of the same equations of motion. 
 
 Once one imposes that a $f(R)$ function has to have Noether symmetries, this translates into mathematical conditions on the $f(R)$ functional form itself. In particular, considering that the class of $f(R)$ functions derived from bosonic string theory has to satisfy these Noether symmetries conditions, one gets a specific functional form for this class.

Finally, one can renormalize the dilaton field by adding a function of the scale factor a(t) of the universe without modifying the equations of motion. As result, one gets a Lagrangian function which is manifestly duality invariant under the transformation on the scale factor of the universe, $a(t)\mapsto 1/a(\pm t)$,  which is the same transformation of the Gasperini-Veneziano pre-Big Bang cosmological models. In this way, from any cosmological solution valid for the scale factor of the universe $a(t)$ one can derive duality symmetric solutions with scale factor of the universe $1/a(\pm t)$. As it is quite clear from the duality transformations, each time one has a cosmological solution valid for positive values of time, it is possible to get a solution for negative values of time as well. Therefore a cosmological model with a starting Big-Bang will have a specular solution that develops before the Big-Bang time and ends in the neighbor of it. 

\section{Conclusions}

In this essay, it has been highlighted that extended theories of gravity are more than "ad hoc" phenomenological theories since, as in same cases treated above, they derive directly from low energy limit of fundamental theories of quantum gravity like string theory. 

Potentially, Extended theories of gravity could shed light in the enigma of understanding the problems of Dark Matter and Dark Energy. This is especially true for those extended theories of gravity that are approximation of Quantum Theory of Gravity. 

The very fact, one has highlighted, that there exist classes of $f(R)$ theories of gravity which are low energy limit of Quantum Gravity Theories is quite important. From the cosmology of these theories it could develop a cosmography, whose agreement with the observational data could discriminate among the different theories of quantum gravity.   

Furthermore, the fact that one has derived the duality symmetry by using Noether symmetry approach suggests to explore the link between the Noether symmetries and duality transformations. The fact that the effective theory of gravity derived from string theory has a manifestly $O(d,d)$ invariance, is questioning how this is linked to the Noether symmetries which have been used to make the class of $f(R)$ Lagrangians manifestly duality invariant.

\end{document}